\documentclass[twocolumn]{article}

\usepackage[english]{babel} \usepackage[utf8x]{inputenc}
\usepackage[T1]{fontenc} \usepackage{authblk}

\usepackage[a4paper,top=3cm,bottom=2cm,left=2cm,right=2cm,marginparwidth=1.5cm]{geometry}

\usepackage{amsmath}
\usepackage{graphicx}
\usepackage[colorinlistoftodos]{todonotes} \usepackage[colorlinks=true,
allcolors=blue]{hyperref}

\usepackage[normalem]{ulem}
\usepackage{amsmath,amsfonts,amssymb}       
\usepackage{bm}
\usepackage{color}
\usepackage{fancyhdr}
\usepackage{cite}
\usepackage{caption}

\DeclareMathAlphabet{\mathbbmsl}{U}{bbm}{m}{sl}
\newcommand\PP{\mathbb{P}}
\newcommand\EE{\mathbb{E}}

\title{CosmoGAN:\@ creating high-fidelity weak lensing convergence maps using Generative Adversarial Networks}
\author[1]{Mustafa Mustafa\thanks{}}
\author[1]{Deborah Bard}
\author[1]{Wahid Bhimji}
\author[1]{Zarija Luki\'c}
\author[2]{Rami Al-Rfou}
\author[3]{Jan M. Kratochvil}
\affil[1]{Lawrence Berkeley National Laboratory, Berkeley, CA 94720}
\affil[2]{Google Research, Mountain View, CA 94043}
\affil[3]{University of KwaZulu-Natal, Westville, Durban 4000, South Africa}

\begin{document}
\date{}
\twocolumn[
	\begin{@twocolumnfalse}
		\maketitle
		\begin{abstract}
      Inferring model parameters from experimental data is a grand challenge
      in many sciences, including cosmology. This often relies critically on
      high fidelity numerical simulations, which are prohibitively
      computationally expensive. The application of deep learning techniques
      to generative modeling is renewing interest in using high dimensional
      density estimators as computationally inexpensive emulators of
      fully-fledged simulations. These generative models have the potential
      to make a dramatic shift in the field of scientific simulations, but for
      that shift to happen we need to study the performance of such generators
      in the precision regime needed for science applications. To this end,
      in this work we apply Generative Adversarial Networks to the problem of
      generating weak lensing convergence maps. We show that our generator
      network produces maps that are described by, with high statistical
      confidence, the same summary statistics as the fully simulated maps.
		\end{abstract}
	\end{@twocolumnfalse}
]
{\renewcommand{\thefootnote}%
    {\fnsymbol{footnote}}
  \footnotetext[1]{Corresponding author: mmustafa@lbl.gov}
}

\section{Introduction}

Cosmology has progressed towards a precision science in the past two
decades, moving from order of magnitude estimates to percent-level
measurements of fundamental cosmological parameters. This was largely
driven by successful CMB and BAO
probes~\cite{Planck2018,Alam2017,Bautista2018,Dumas2017}, which
extract information from very large scales, where structure formation
is well described by linear theory or by the perturbation
theory~\cite{Carlson2009}. In order to resolve the next set of
outstanding problems in cosmology---for example the nature of dark
matter and dark energy, the total mass and number of neutrino species,
and primordial fluctuations seeded by inflation---cosmologists will
have to rely on measurements of cosmic structure at far smaller scales.

Modeling the growth of structure at those scales involves non-linear
physics that cannot be accurately described analytically, and we instead
use numerical simulations to produce theoretical predictions which are
to be confronted with observations. Weak gravitational lensing is
considered to be one of the most powerful tools of extracting
information from small scales. It probes both the geometry of the
Universe and the growth of structure~\cite{Bartelmann2001}. The
shearing and magnification of background luminous sources by
gravitational lensing allows us to reconstruct the matter distribution
along the line of sight. Common characterizations for gravitational
lensing shear involve cross-correlating the ellipticities of galaxies
in a two-point function estimator, giving the lensing power spectrum.
By comparing such measurements to theoretical predictions, we can
distinguish whether, for example, cosmic acceleration is caused by dark
energy or a modification of general relativity.

The scientific success of the next generation of photometric sky surveys
(e.g.~\cite{euclid}, \cite{lsst}, \cite{wfirst}) therefore hinges critically on the
success of underlying simulations. Currently the creation of each
simulated virtual universe requires an extremely computationally
expensive simulation on High Performance Computing (HPC) resources. That
makes direct application of Markov chain Monte Carlo
(MCMC,~\cite{Metropolis1953}, \cite{Gelman2004}) or similar Bayesian methods
prohibitively expensive, as they require hundreds of thousands of
forward model evaluations to determine the posterior probabilities of
model parameters. In order to make this inverse problem practically
solvable, constructing a computationally cheap \emph{surrogate model}
or an \emph{emulator}~\cite{coyote2,coyote3} is imperative. However,
traditional approaches to emulators require the use of the summary
statistic which is to be emulated. An approach that makes no assumptions
about such mathematical templates of the simulation outcome would be of
considerable value. While in this work we focus our attention on the
generation of the weak lensing convergence maps, we believe that the
method presented here is relevant to many similar problems in
astrophysics and cosmology where a large number of expensive simulations
is necessary.

Recent developments in deep generative modeling techniques open the
potential to meet this emulation need. The density estimators in these
models are built out of neural networks which can serve as universal
approximators~\cite{universalapproximator}, thus having the ability
to learn the underlying distributions of data and emulate the observable
without imposing the choice of summary statistics, as in the traditional
approach to emulators. These data-driven generative models have also
found applications in astrophysics and cosmology. On the observational
side, these models can be used to improve images of galaxies beyond the
deconvolution limit of the telescope~\cite{Schawinski2017}. On the
simulation side, they have been used to produce samples of cosmic
web~\cite{webgan}.

In addition to generative modeling, we have recently witnessed a number
of different applications of deep learning and convolutional neural
networks (CNN) to the problem of parameter inference using weak
gravitational lensing. Training deep neural networks to regress
parameters from data simulations~\cite{Gupta2018,Ribli2019} shows
that CNN trained on noiseless lensing maps can bring few times tighter
constraints on cosmological parameters than the power spectrum or
lensing peaks analysis methods. Similar analysis has been performed for
the case of simulated lensing maps with varying level of noise due to
imperfect measurement of galaxy shape distortions and finite number
density of the source galaxies~\cite{Fluri2018}. Although the
advantage of CNNs over ``traditional'' methods is still present in the
noisy case, the quantitative improvement is much less than in the case
of noiseless simulated maps. CNNs applied to weak lensing maps have also
been proposed for improved differentiating between dark energy and
modified gravity cosmologies~\cite{Peel2018}. Finally, an
image-to-image translation method employing conditional adversarial
networks was used to demonstrate learning the mapping from an input,
simulated noisy lensing map to the underlying noise
field~\cite{Shirasaki2018}, effectively denoising the map.

We would like to caution that all deep learning methods are very
sensitive to the input noise, thus networks trained on either noiseless
or even on simulated, idealized, noise are likely to perform poorly in
the inference regime on the real data without a complex and
domain-dependent preparation for such task~\cite{domainadaptation}.
This problem is exaggerated by the fact that trained CNNs are mapping
input lensing map to cosmological parameters without providing full
posterior probability distribution for these parameters or any other
good description of inference errors.

In this work, we study the ability of a variant of generative models,
Generative Adversarial Networks (GANs)~\cite{gan} to generate weak
lensing convergence maps. In this paper, we are not concerned about
inferring cosmological parameters, and we do not attempt to answer
questions of optimal parameter estimators, either with deep learning or
``traditional'' statistical methods. Here, we study the ability of GANs
to produce convergence maps that are statistically indistinguishable from maps
produced by physics-based generative models, which are in this case
N-body simulations. The training and validation maps are produced using
N-body simulations of $\varLambda $CDM cosmology. We show that maps
generated by the neural network exhibit, with high statistical
confidence, the same power spectrum of the fully-fledged simulator maps,
as well as higher order non-Gaussian statistics, thus demonstrating that
such scientific data can be amenable to a GAN treatment for generation.
The very high level of agreement achieved offers an important step
towards building emulators out of deep neural networks.

The paper is organized as follows: in Sect.~\ref{sec:methods} we
outline the data set used and describe our GAN architecture. We present
our results in Sect.~\ref{sec:Results} and we outline the future
investigations which we think are critical to build weak lensing
emulators in Sect.~\ref{sec:discussion}. Finally, we present
conclusions of this paper in Sect.~\ref{sec:conclusion}.

\begin{figure*}[ht] \centering
  \includegraphics[width=0.95\textwidth]{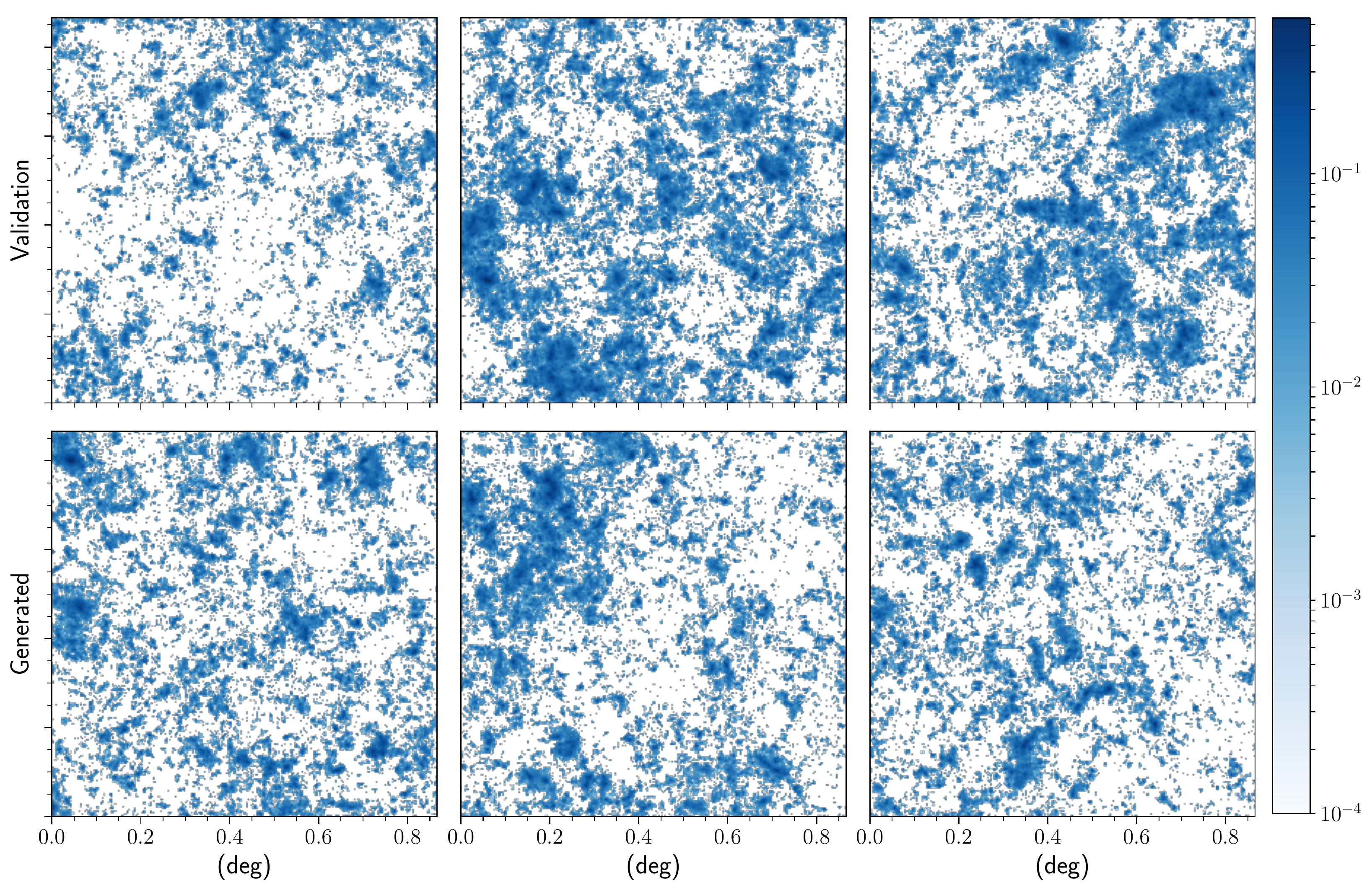}
  \caption{\label{fig:generated} Weak lensing convergence maps for our
  $\Lambda$CDM cosmological model. Randomly selected maps from
  validation dataset (top) and GAN generated examples (bottom).}
\end{figure*}

\section{Methods}\label{sec:methods}
In this section we first introduce the weak lensing convergence maps used in
this work, and then proceed to describe Generative Adversarial Networks and
their implementation in this work.

\subsection{Dataset}\label{sec:dataset}
To produce our training dataset, we use the cosmological simulations
described in~\cite{Kratochvil-MF-sims}, \cite{Yang:2011zzn}, produced using
the Gadget2~\cite{Gadget2} N-Body simulation code and ray-traced with
the Inspector Gadget weak lensing simulation
pipeline~\cite{Kratochvil-MF-sims,Kratochvil:2009wh,Yang:2011zzn} to
produce weak lensing shear and convergence maps. A total of 45
simulations were produced, each consisting of 512$^{3}$ particles in a
box of size of 240~$h^{-1}$Mpc. The cosmological parameters used in these
simulations were {$\sigma _{8} = 0.798$, $w=-1.0$, $\varOmega _{m} = 0.26$,
$\varOmega _{\varLambda }=0.74$, $n_{s} = 0.96$, $H_{0} = 0.72$}. These
simulation boxes were rotated and translated to produce 1000 ray-traced
lensing maps at the redshift plane $z= 1.0$. Each map covers 12 square
degrees, with $2048\times 2048$ pixels, which we downsampled to
$1024\times 1024$ pixels. Following the formalism introduced
in~\cite{BJ}, the gravitational lensing illustrated by these maps can
be described by the Jacobian matrix
\begin{equation}
A(\theta) =
\begin{bmatrix} 1-\kappa - \gamma_1 & -\gamma_2 \\
-\gamma_2 & 1-\kappa+\gamma_1
\end{bmatrix}
\end{equation}
where $\kappa$ is the convergence, and $\gamma$ is the shear.

The training data was generated by randomly cropping $200$ $256\times
256$ maps from each of the original $1000$ maps. The validation and
development datasets have been randomly cropped in the same manner. We
have tested sampling the validation datasets from \textit{all of} the
original $1000$ maps versus sampling the training from $800$ maps and
the validation from the remaining $200$ maps. GANs are not generally
amenable to memorization of the training dataset, this is in part
because the generator network isn't trained directly on that data; it
only learns about it by means of information from another network, the
discriminator, as is described in the next section. Therefore, in our
studies, it did not make any difference how we sample our validation
dataset. We demonstrate that the generator network did not memorize the
training dataset in Sect.~\ref{sec:generalize}. Finally, the
probability for a map to have a single pixel value outside [$-1.0,1.0$]
range is less than $0.9\%$ so it was safe to use the data without any
normalization.

In one of the tests we report in this paper we use an auxiliary dataset,
which consists of 1000 maps produced using the same simulation code and
cosmological parameters, but with a different random seed, resulting in
a set of convergence maps that are statistically independent from those
used in our training and validation.

\subsection{Generative Adversarial Networks}
The central problem of generative models is the question: given a
distribution of data $\mathbb{P}_{r}$ can one devise a generator $G$
such that the distribution of model generated data $\mathbb{P}_{g} =
\mathbb{P}_{r}$? Our information about $\mathbb{P}_{r}$ comes from the
training dataset, typically an independent and identically distributed
random sample $x_{1},x_{2}, \ldots, x_{n}$ which is assumed to have the
same distribution as $\mathbb{P}_{r}$. Essentially, a generative model
aims to construct a density estimator of the dataset. The GAN frameworks
constructs an implicit density estimator which can be efficiently
sampled to generate samples of $\mathbb{P}_{g}$.

The GAN framework~\cite{gan} sets up a game between two players, a
generator and a discriminator. The generator is trained to generate
samples that aim to be indistinguishable from training data as judged
by a competent discriminator. The discriminator is trained to judge
whether a sample looks real or fake. Essentially, the generator tries
to fool the discriminator into judging a generated map looks real.

In the neural network formulation of this framework the generator
network $G_{\phi }$, parametrized by network parameters $\phi $, and
discriminator network $D_{\theta }$, parametrized by $\theta $, are
simultaneously optimized using gradient descent. The discriminator is
trained in a supervised manner by showing it real and generated samples,
it outputs a probability of the input map being real or fake. It is
trained to minimize the following cross-entropy cost function:

\begin{equation}
    \label{eq:game_loss}
    J^{(D)} = -\EE_{x\sim\PP_{r}} \log{D_{\theta}(x)} - \EE_{x\sim\PP_{g}} \log{(1 - D_{\theta}(x)).}
\end{equation}

The generator is a differentiable function (except at possibly finitely
many points) that maps a sample from a noise prior, $\boldsymbol{z} \sim p(
\boldsymbol{z})$, to the support of $\mathbb{P}_{g}$. For example, in this work,
the noise vector is sampled from a 64-dimensional isotropic normal
distribution and the output of the generator are maps $x \in
\mathbb{R}^{256\times 256}$. The dimension of the noise vector
$\boldsymbol{z}$ needs to be commensurate with the support of the convergence
maps $\mathbb{P}_{r}$ in $\mathbb{R}^{256\times 256}$. In the
game-theoretic formulation, the generator is trained to maximize
equation~\eqref{eq:game_loss}, this is known as the minimax game. However,
in that formulation, the gradients of the cost function with respect to
the generator parameters vanish when the discriminator is winning the
game, i.e. rejecting the fake samples confidently. Losing the gradient
signal makes it difficult to train the generator using gradient descent.
The original GAN paper~\cite{gan} proposes flipping the target for the
generator instead:

\begin{equation}
    \label{eq:heuristic}
    J^{(G)} = - \EE_{x\sim\PP_{g}} \log{D_{\theta}(x)}. 
\end{equation}

This ``heuristically'' motivated cost function (also known as the
non-saturating game) provides strong gradients to train the generator,
especially when the generator is losing the game~\cite{gantutorial}.

Since the inception of the first GAN, there have been many proposals for
other cost functions and functional constrains on the discriminator and
generator networks. We have experimented with some of these but in
common with a recent large scale empirical study~\cite{largeganstudy}
of these different models: ``did not find evidence that any of the
tested algorithms consistently outperforms the non-saturating GAN'',
introduced in~\cite{gan} and outlined above. That study attributes the
improvements in performance reported in recent literature to difference
in computational budget. With the GAN framework laid out we move to
describing the generator and discriminator networks and their training
procedure.

\begin{figure}[ht] \centering
  \includegraphics[width=0.43\textwidth]{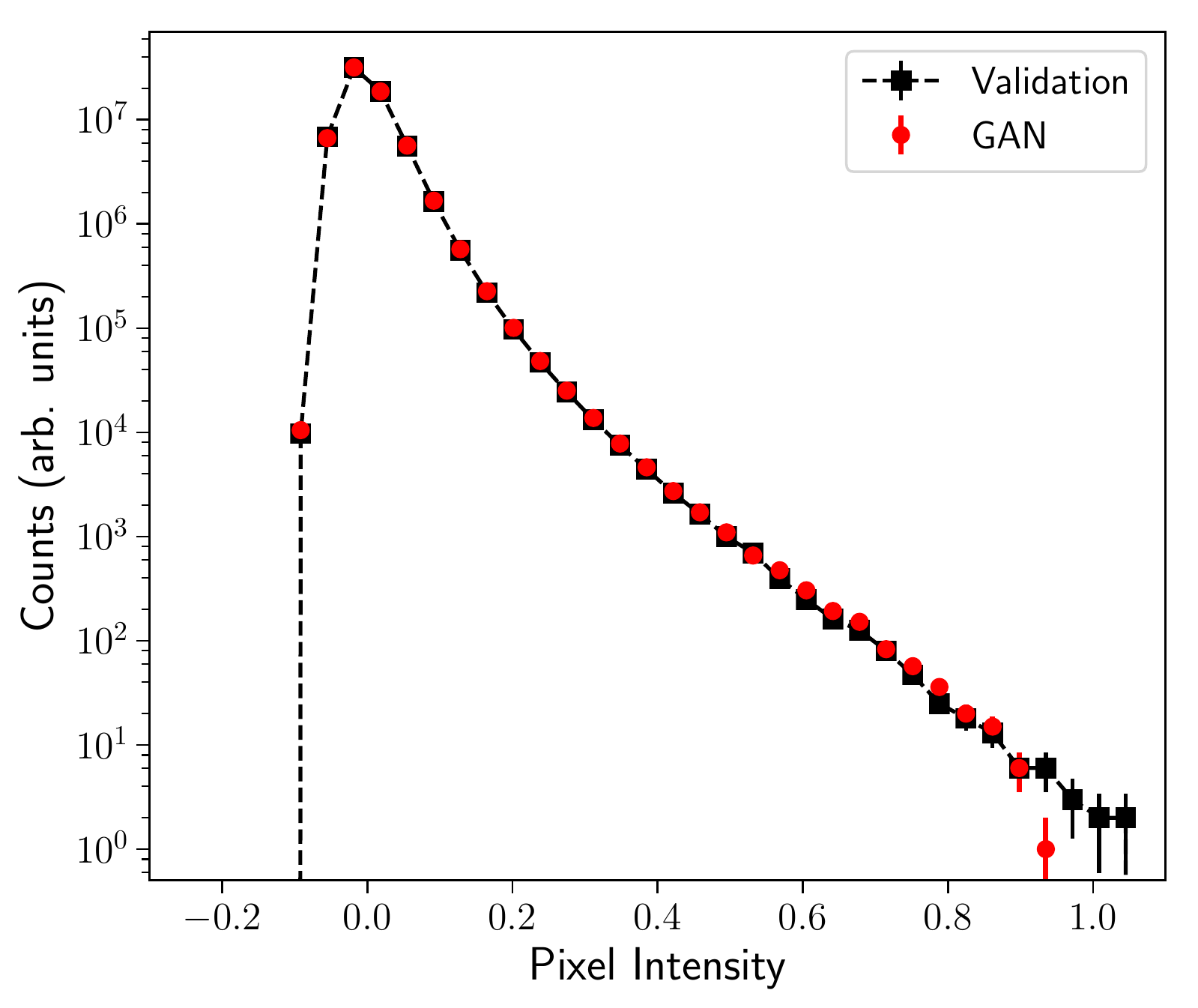}
  \caption{\label{fig:pixel_intensity}Pixel intensity distribution of 1000
  generated maps (red circles) compared to those of 1000 validation maps (black
  squares). The GAN is able to emulate the distribution of intensities in the
  maps. The Kolmogorov-Smirnov test of similarity of these distributions gives
  a p-value $>0.999$.}
\end{figure}

\subsection{Network architecture and training}\label{sec:arch}

\begin{figure*}[ht] \centering
 \includegraphics[width=0.86\textwidth]{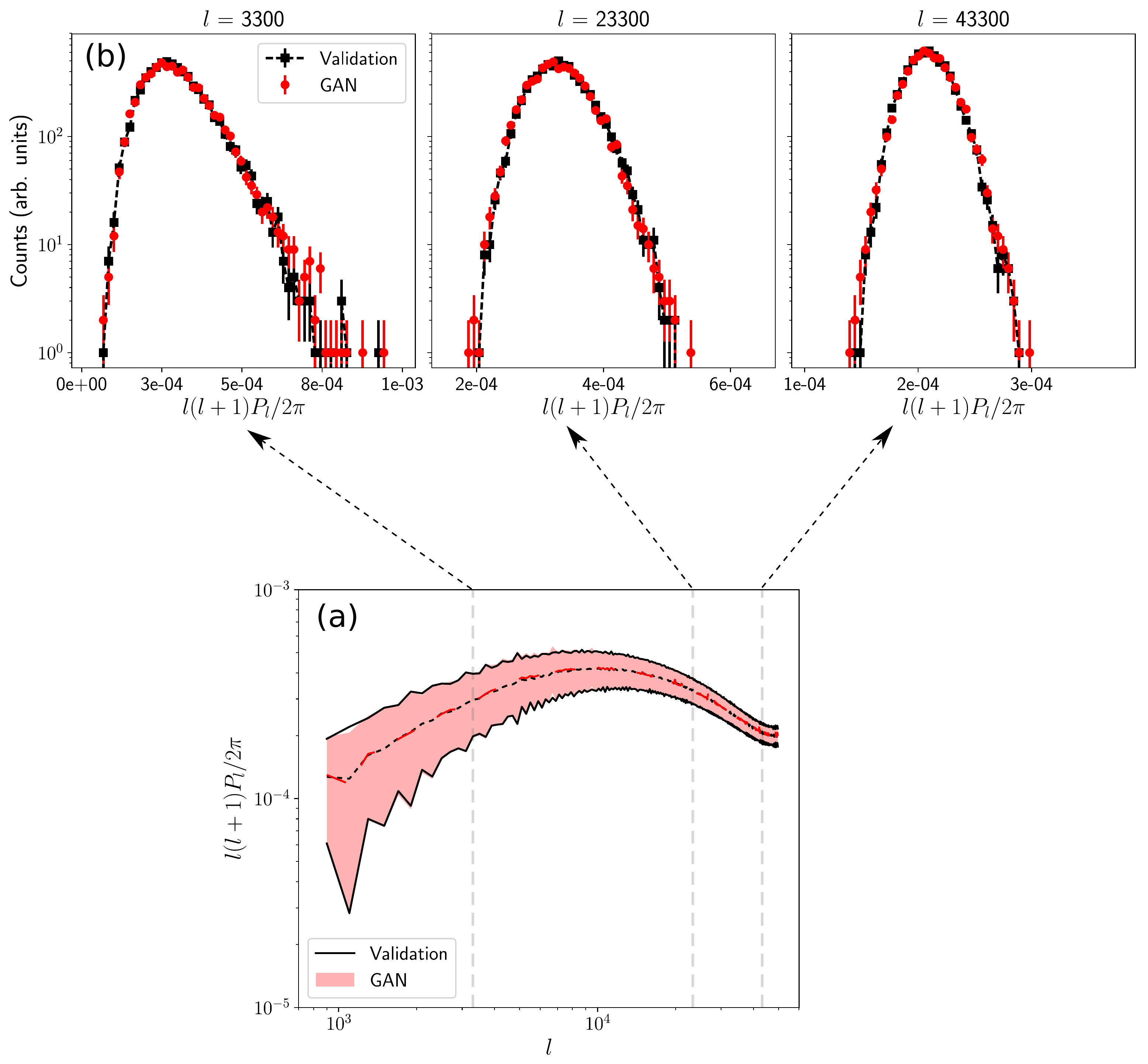}
 \caption{\label{fig:ps}The power spectrum of the convergence maps, evaluated
  at 248 Fourier modes. We use 100 batches of generated maps (6400 in total)
  for this comparison. Shown in (a) are bands of the $\mu(l)\pm\sigma(l)$, the
  dashed lines represent the means $\mu(l)$, (b) shows the underlying
  distributions at 3 equidistant modes for illustration. }
\end{figure*}

Given the intrinsic translation invariance of the convergence
maps~\cite{cosmoreview}, it is natural to construct the generator and
discriminator networks mainly from convolutional layers. To allow the
network to learn the proper correlations on the components of the input
noise $\boldsymbol{z}$ early on, the first layer of the generator network needs
to be a fully-connected layer. A class of all convolutional network
architectures has been developed in~\cite{allconv}, which use strided
convolutions to downsample instead of pooling layers, and also use
strided transposed convolutions to upsample. This architecture was later
adapted to build GANs in Deep Convolutional Generative Adversarial
Networks (DCGAN)~\cite{Radford2015}. We experimented with DCGAN
architectural parameters and we found that most of the hyper-parameters
optimized for natural images by the original authors perform well on the
convergence maps. We used DCGAN architecture with slight modifications
to meet our problem dimensions, we also halved the number of filters.\looseness=1

The generator takes a 64-dimensional vector sampled from a normal prior
$z\sim \mathcal{N}(0,1)$. The first layer is a fully connected layer
whose output is reshaped into a stack of feature maps. The rest of the
network consists of four layers of transposed convolutions (a convolutional layer
with fractional strides where zeroes are inserted between each column
and row of pixels before convolving the image with the filter in order
to effectively up-sample the image) that lead to a single channel
$256\times 256$ image. The outputs of all layers, except the output one,
are batch-normalized~\cite{batchnorm} (by subtracting the mean
activation and dividing by its standard deviation and learning a linear
scaling) which was found to stabilize the training. A~rectified linear
unit (ReLU) activation~\cite{relu} (output zero when the input less
than zero and output equal to the input otherwise) is used for all
except the output layer where a hyperbolic-tangent ($\tanh $) is used.
The final generator network architecture is summarized in
Table~\ref{table:generator}.

\begin{table}[ht]
\begin{tabular}{lccc}
\hline
 & \textbf{Activ.} & \textbf{Output shape}  & \textbf{Params.} \\ \hline
Latent             & -               & 64                     & -                \\ \hline
Dense              & -               & 512 $\times$ 16 $\times$ 16  & 8.5M             \\
BatchNorm          & ReLU            & 512 $\times$ 16 $\times$ 16  & 1024             \\ \hline
TConv 5 $\times$ 5    & -               & 256 $\times$ 32 $\times$ 32  & 3.3M             \\
BatchNorm          & ReLU            & 256 $\times$ 32 $\times$ 32  & 512              \\ \hline
TConv 5 $\times$ 5    & -               & 128 $\times$ 64 $\times$ 64  & 819K             \\
BatchNorm          & ReLU            & 128 $\times$ 64 $\times$ 64  & 256              \\ \hline
TConv 5 $\times$ 5    & -               & 64 $\times$ 128 $\times$ 128 & 205K             \\
BatchNorm          & ReLU            & 64 $\times$ 128 $\times$ 128 & 128              \\ \hline
TConv 5 $\times$ 5    & Tanh            & 1 $\times$ 256 $\times$ 256  & 1601             \\ \hline
\multicolumn{3}{l}{Total trainable parameters}                & \textbf{12.3M}   \\ \hline
\end{tabular}
\caption{Generator network architecture: layer types, activations, output
shapes (channels $\times$ height $\times$ width) and number of trainable
parameters for each layer. TConv are Transposed Convolution layers with
strides$=2$.}\label{table:generator}
\end{table}

\begin{table}[ht]
\begin{tabular}{lccc}
\hline
 & \textbf{Activ.} & \textbf{Output shape}  & \textbf{Params.} \\ \hline
Input map              & -               & 1 $\times$ 256 $\times$ 256  & -                \\ \hline
Conv 5 $\times$ 5         & LReLU           & 64 $\times$ 128 $\times$ 128 & 1664             \\ \hline
Conv 5 $\times$ 5         & -               & 128 $\times$ 64 $\times$ 64  & 205K             \\
BatchNorm              & LReLU           & 128 $\times$ 64 $\times$ 64  & 256              \\ \hline
Conv 5 $\times$ 5         & -               & 256 $\times$ 32 $\times$ 32  & 819K             \\
BatchNorm              & LReLU           & 256 $\times$ 32 $\times$ 32  & 512              \\ \hline
Conv 5 $\times$ 5         & -               & 512 $\times$ 16 $\times$ 16  & 3.3M             \\
BatchNorm              & LReLU           & 512 $\times$ 16 $\times$ 16  & 1024             \\ \hline
Linear                 & Sigmoid         & 1                      & 131K             \\ \hline
\multicolumn{3}{l}{Total trainable parameters}                    & \textbf{4.4M}    \\ \hline
\end{tabular}
\caption{Discriminator network architecture: layer types, activations, output
shapes (channels $\times$ height $\times$ width) and number of trainable
parameters for each layer. All convolutional layers have stride $=2$.
LeakyReLU's leakines $=0.2$.}\label{table:disciminator}
\end{table}

\begin{figure*}[ht] \centering
 \includegraphics[width=0.98\textwidth]{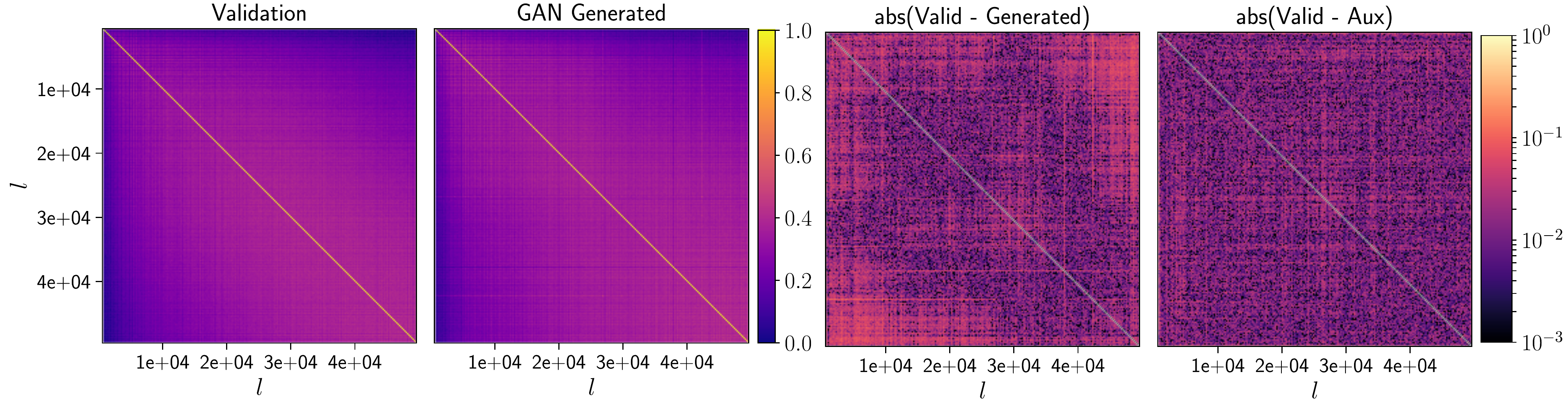}
  \caption{\label{fig:ps_corr} The correlation matrices of the modes of the
  power spectra shown in Fig.~\ref{fig:ps}. The first two panels show the
  correlation matrices of the validation and the GAN generated maps, the third
  panel shows the difference between these correlation matrices. To provide a
  scale for comparison, in the fourth panel we also show the difference between
  the validation dataset correlation matrix and the correlations in an
  auxiliary dataset (see text for details).}
\end{figure*}

The discriminator has four convolutional layers. The number of feature
maps, stride and kernel sizes are the same as in the generator. We
reduced the number of filters from the DCGAN guidelines by a factor of
2, which effectively reduces the capacity of the
generator/discriminator. This worked well for our problem and reduced
the training time. The output of all convolutional layers are activated
with LeakyReLU~\cite{lrelu} with parameter $\alpha =0.2$. The output
of the last convolutional layer is flattened and fed into a fully
connected layer with a 1-dimensional output that is fed into a sigmoid.
Batch Normalization is applied before activations for all layers'
outputs except the first layer. The final discriminator network
architecture is summarized in Table~\ref{table:disciminator}.

Finally, we minimize discriminator loss in Eq.~\eqref{eq:game_loss} and
generator loss in Eq.~\eqref{eq:heuristic} using Adam
optimizers~\cite{adam} with the parameters suggested in the DCGAN
paper: learning rate $0.0002$ and $\beta _{1}=0.5$. Batch-size is 64
maps. We flip the real and fake labels with $1\%$ probability to avoid
the discriminator overpowering the generator too early into the
training. We implement the networks in TensorFlow~\cite{tf} and train
them on a single NVIDIA Titan X Pascal GPU.

Training GANs using a heuristic loss function often suffers from
unstable updates towards the end of their training. This has been
analyzed theoretically in~\cite{arjovsky2017a} and shown to happen
when the discriminator is close to optimality but has not yet converged.
In other words, the precision of the generator at the point we stop the
training is completely arbitrary and the loss function is not useful to
use for early stopping. For the results shown in this work we trained
the network until the generated maps pass the visual and pixel intensity
Kolmogorov--Smirnov tests, see Sect.~\ref{sec:Results}. This took 45
passes (epochs) over all of the training data. Given the un-stability
of the updates at this point, the performance of the generator on the
summary statistics tests starts varying uncontrollably. Therefore, to
choose a generator configuration, we train the networks for two extra
epochs after epoch 45, and randomly generate 100 batches (6400 maps) of
samples at every single training step. We evaluate the power spectrum
on the generated samples and calculate the Chi-squared distance
measurement to the power spectrum histograms evaluated on a development
subset of the validation dataset. We use the generator with the best
Chi-squared distance.

\begin{figure*}[t] \centering
  \includegraphics[width=\textwidth]{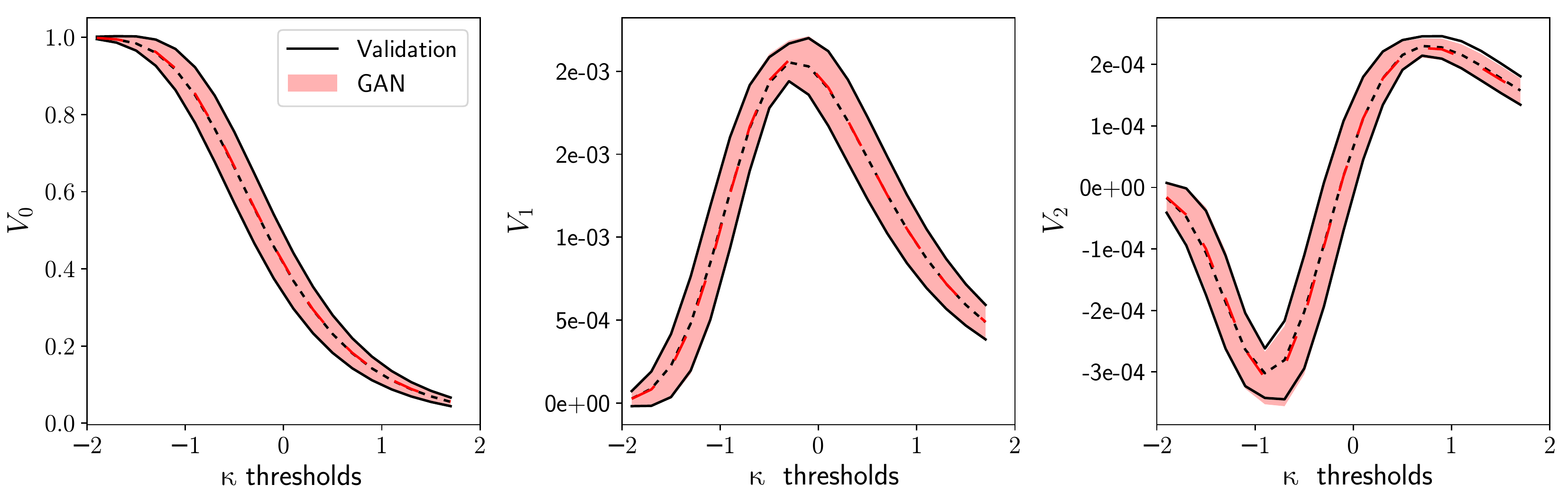}
  \caption{\label{fig:minkowski_bands}The Minkowski functionals of the
  convergence maps, which are sensitive to the non-Gaussian structures. We
  carried out the measurements on 100 batches of generated maps (6400 in total)
  and compare them to those of the validation maps. The functionals are
  evaluated at 19 thresholds and shown here are the bands of $\mu\pm\sigma$ at
  each threshold. The dashed lines represent the mean $\mu$.}
\end{figure*}

\begin{figure*}[ht] \centering
  \includegraphics[width=0.6\textwidth]{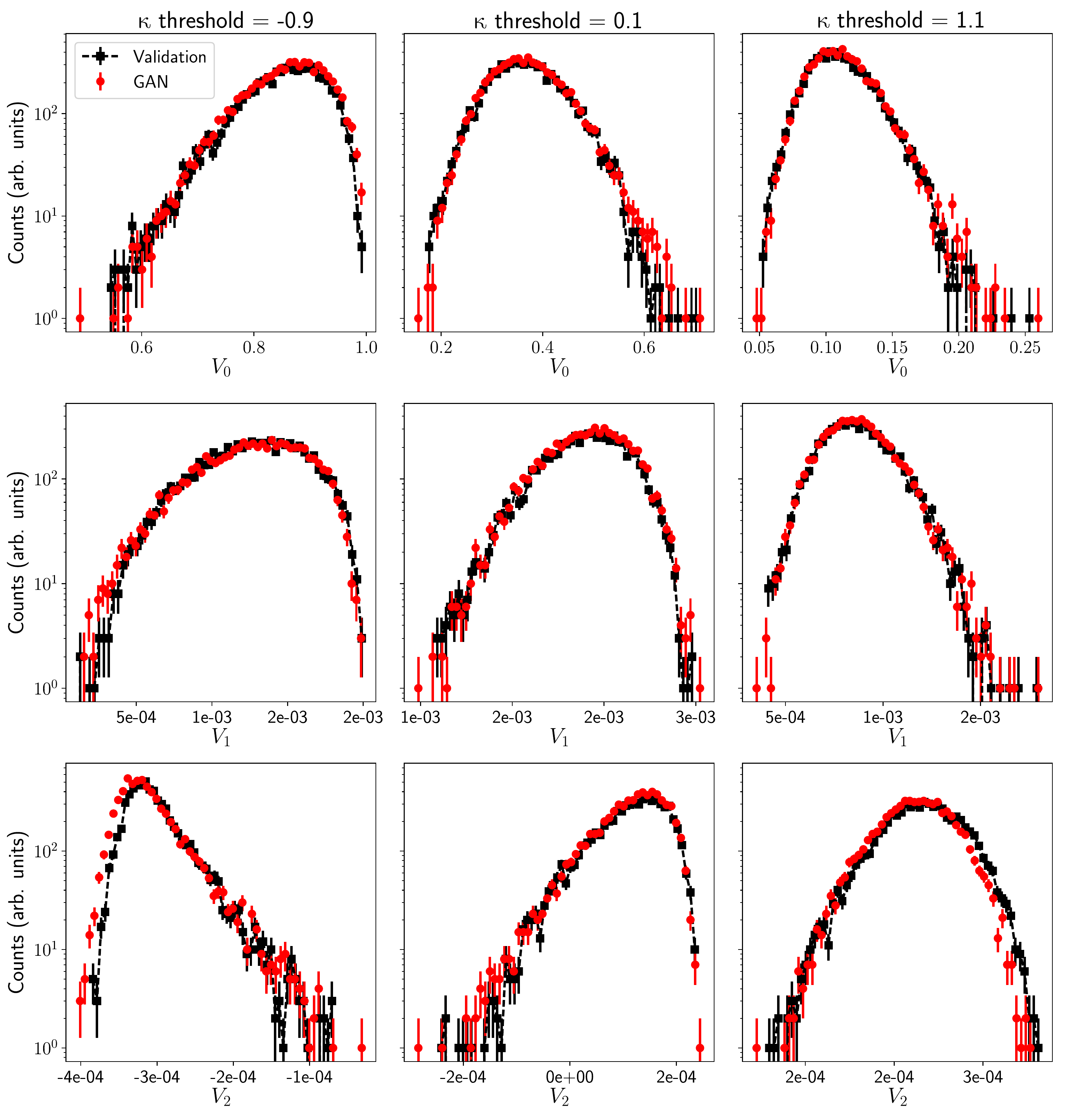}
   \caption{\label{fig:minkowski_slices}The distributions of the Minkowski
  functionals at 3 equidistant thresholds shown for illustration.}
\end{figure*}

\section{Results}\label{sec:Results}

Figure~\ref{fig:generated} shows examples of maps from the validation
and GAN generated datasets. The validation dataset has not been used in
the training or tuning of the networks. Visually, an expert cannot
distinguish the generated maps from the full simulation ones.

\subsection{Evaluation of generator fidelity}

Once we have obtained a density estimator of the original data the first
practical question is to determine the goodness of the fit. Basically,
\textit{how close is $\mathbb{P}_{g}$ to $\mathbb{P}_{r}$?} This issue
is critical to understanding and improving the formulation and training
procedure for generative models, and is an active area of
research~\cite{theis2015}. Significant insight into the training
dynamics of GANs from a theoretical point of view has been gained
in~\cite{arjovsky2017a}, \cite{wgan} and later works. We think that when it
comes to practical applications of generative models, such as in the
case of emulating scientific data, the way to evaluate generative models
is to study their ability to reproduce the charachtarestic statistic of the original dataset.

To this end, we calculate three statistics on the generated convergence
maps: a first order statistic (pixel intensity), the power spectrum and
a non-Gaussian statistic. The ability to reproduce such summary
statistics is a reliable metric to evaluate generative models from an
information encoding point of view. To test our statistical confidence
of the matching of the summary statistics we perform bootstrapped
two-tailed Kolmogorov--Smirnov (KS) test and Andersen--Darling (AD) test
of the null-hypothesis that the summary statistics in the generated maps
and the validation maps have been drawn from the same continuous
distributions~.\footnote{We used a ROOT~\cite{ROOT} implementation
of the Andersen--Darling test and Scipy~\cite{scipy} for the
Kolmogorov--Smirnov test.}

Figure~\ref{fig:pixel_intensity} shows a histogram of the distribution
of pixel intensities of an ensemble of generated maps compared to that
of a validation dataset. It is clear that the GAN generator has been
able to learn the probabilities of pixel intensities in the real
simulation dataset, the KS p-value is $>0.999$. We note that the maps
generated by this GAN have the same the geometry of the simulated maps,
i.e. angular size, resolution, etc.

The second-order statistical moment in gravitational lensing is the
shear power spectrum. This is a measure of the correlation in
gravitational lensing at different distances, and characterizes the
matter density fluctuations at different length scales. Assuming we have
only Gaussian fluctuations $\delta (x)$ at comoving distance $x$, the
matter density of the universe can be defined by its power spectrum
$P_{\kappa }$:

\begin{equation} 
  \langle \tilde{\delta}(l) \tilde{\delta}^{*}(l') \rangle = (2\pi)^{2} \delta_{D}(l-l') P_{\kappa}(l)
\end{equation}
where $\delta _{D}$ is the Dirac delta function, and $l$ is the Fourier
mode~\cite{cosmoreview}. The power spectrum (and its corresponding
real-space measurement, the two-point correlation function) of
convergence maps has long been used to constrain models of cosmology by
comparing simulated maps to real data from sky
surveys~\cite{CFHTLenS-PS,DES-PS,DLS-PS}. Numerically, the power
spectrum is the amplitudes of the Fourier components of the map. We
calculate the power spectrum at 248 Fourier modes of an ensemble of
generated maps using LensTools~\cite{lenstools}, and compare them to
the validation dataset. Since each map is a different random
realization, the comparison has to be made at the ensemble level.
Figure~\ref{fig:ps}(a) shows two bands representing the mean
$\mu (l)$ $\pm $ standard deviation $\sigma (l)$ of the ensemble of
power spectra at each Fourier mode of the validation and a generated
dataset. As is clear in the figure, the bands completely overlap with
each other. To confirm that the range of the power spectrum at a given
$l$ is completely reproduced in the generated maps we look
differentially at the underlying 1-D distributions. Samples of such
distributions at equally spaced Fourier modes are shown in
Fig.~\ref{fig:ps}(b). The bootstrapped KS (AD) p-values for $236$
($205$) Fourier modes are $>0.99$, of the remaining modes, $10$ ($26$)
are $>0.95$, all the remaining are larger than $0.8$. The power spectrum
is the figure of merit for evaluating the success of an emulator for
reproducing the structures of the convergence map, and we have shown
with statistical confidence that GANs are able to discover and reproduce
such structures. It should be noted that the power spectra of the
generated maps is limited to the scale resolutions in the training
dataset. For example in Fig.~\ref{fig:ps}(a), one can see the
statistical fluctuations at the lower modes in the validation dataset,
the generator faithfully reproduces those fluctuations.

\begin{figure*}[ht] \centering
  \includegraphics[width=\textwidth]{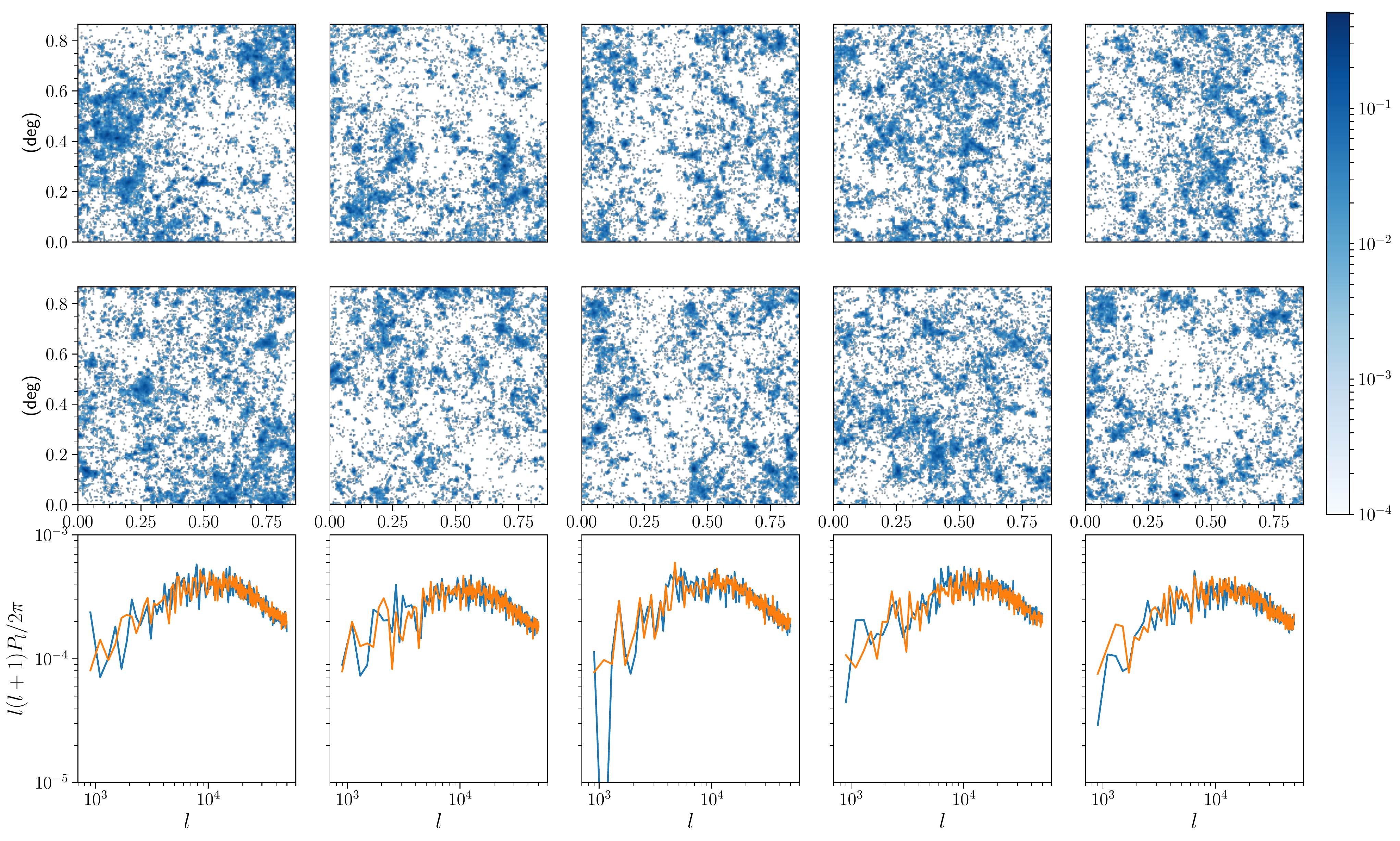}
  \caption{\label{fig:close} Comparison of randomly generated maps (top) with
  training maps (middle) that are their nearest-neighbor in terms of L$^2$
  distance in power spectrum. While the power spectra (bottom) match very well,
  the maps are clearly different.}
\end{figure*}
In Fig.~\ref{fig:ps_corr} we show the correlation matrices of the power
spectra shown in Fig.~\ref{fig:ps}, and the difference between the
correlation matrices from the validation maps and those produced by our
GAN. We also show a comparison of the correlation matrices from the
validation dataset and an the auxiliary dataset (a statistically
independent dataset generated using the procedure used for the original
dataset but with a different random seed, see~Sect.~\ref{sec:dataset}).

There are differences between the validation and auxiliary correlations
matrices of up to 7\%. The differences between the validation and
generated correlation matrices are of a similar level (up to 10\%) but
show a pattern of slightly higher correlation between high- and low-$l$
modes. We find it interesting that the GAN generator assumes a slightly
stronger correlation between small and large modes, thus a smoother
power spectra, than in the simulations. This is due to the large
uncertainty of the power spectra at small modes in the original
simulations as seen in Fig.~\ref{fig:ps}(a).

The power spectrum only captures the Gaussian components of matter
structures in the universe. However, gravity produces non-Gaussian
structures at small scales which are not described by the two-point
correlation function~\cite{cosmoreview}. There are many ways to access
this non-Gaussian statistical information, including higher-order
correlation functions, and topological features such as Minkowski
functionals $V_{0}$, $V_{1}$, and
$V_{2}$~\cite{minkowski,CFHTLenS-MF,Kratochvil-MF-sims} which
characterize the area, perimeter and genus of the convergence maps. We
investigate the ability of the networks to discover and reproduce these
non-Gaussian structures in the maps by evaluating the three Minkowski
functionals. Figure~\ref{fig:minkowski_bands} compares bands of
$\mu \pm \sigma $ of the three Minkowski Functionals (calculated
using~\cite{lenstools}) to those in the real dataset. Each functional
is evaluated at 19 thresholds. As we did with the power spectrum, we
show the Minkowski functionals at different thresholds in
Fig.~\ref{fig:minkowski_slices}. The bootstrapped KS (AD) for $40$
($32$) threshold histograms are $>0.95$ p-values $7$ ($6$) are
$>0.9$ all the remaining are $>0.6$. Again, it is clear that the GANs
have successfully reproduced those non-Gaussian structures in the maps.

Finally, we measure the approximate speed of generation:, it takes the
generator $\approx $15~s to produce 1024 images on a single NVIDIA Titan
X Pascal GPU. Training the network takes $\approx $4 days on the same
GPU. Running N-Body simulations of a similar scale used in this paper
requires roughly 5000 CPU hours per box, with an additional 100-500 CPU
hours to produce the planes and do the ray-tracing to make the 1000 maps
per box. While this demonstrates that, as expected, GANs offer
substantially faster generation, it should be cautioned that these
comparisons are not ``apple-to-apple'' as the N-Body simulations are
ab-initio, while the GANs generated maps are by means of sampling a fit
to the data.

\subsection{Is the generator creating novel maps?}\label{sec:generalize}

\begin{figure*}[ht] \centering
  \includegraphics[width=\textwidth]{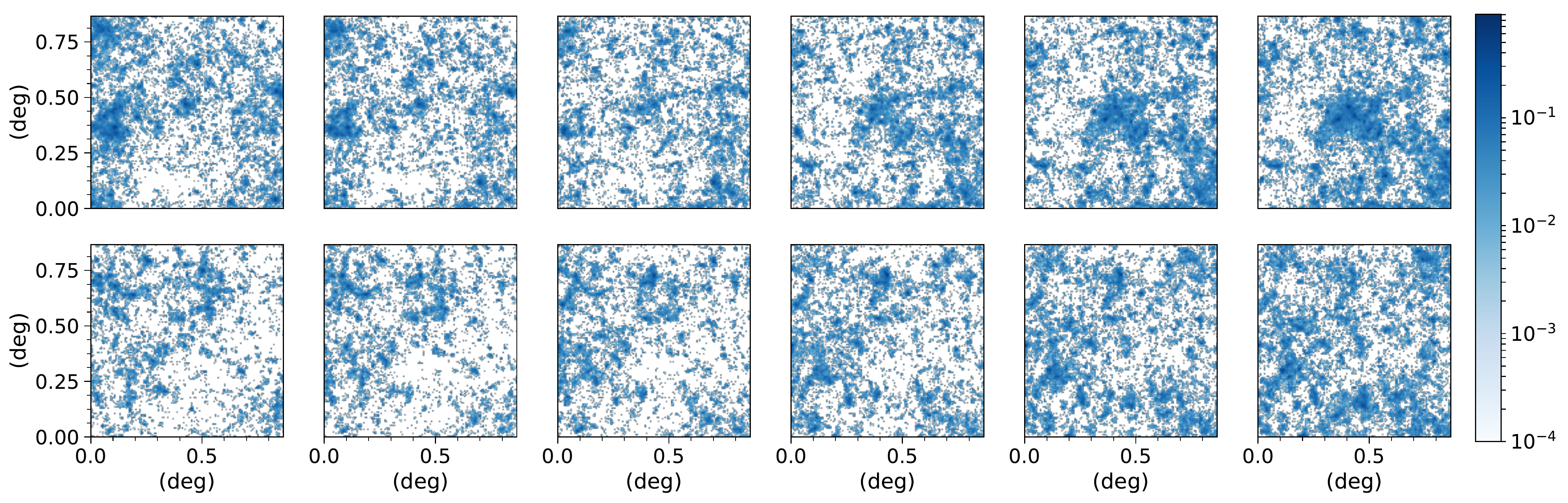}
  \caption{\label{fig:zinterploate} Each row is a random example of maps
produced when interpolating between two randomly chosen vectors $\bm{z}$ (left
to right). The generator varies smoothly between points in the prior space.}
\end{figure*}
The success of the generator at creating maps that matches the summary
statistics of the validation dataset raises the question of whether the
generator is simply memorizing the training dataset. Note that in the
GAN framework, the generator never sees the training dataset directly.
The question is if the generator learned about the training dataset
through the gradients it gets from the discrminator. We conduct two
tests to investigate this possibility.

In the first test we generate a set of maps and find the training maps
that are closest to them. We test two distance notions, the Euclidean
distance (L$^{2}$) in the pixel space of the maps and in the power
spectrum space. Some examples of the latter are shown in
Fig.~\ref{fig:close}. Using both metrics, the training data points that
are closest to generated maps are visually distinct. Concluding that the
generator is producing maps that are not present in the training dataset.

In the second test we investigate whether the generator is a smooth
function on $\boldsymbol{z}$. Essentially, does the generator randomly associate
points in the prior space with maps or, learns a smooth and dense
mapping?. This can be tested by randomly choosing two noise vectors and
evaluating the generator on the line connecting them. We use spherical
linear interpolation to interpolate in the Gaussian prior
space~\cite{slerp}. Figure~\ref{fig:zinterploate} shows examples from
this test. It is clear that the generated maps smoothly morph from one
to another when the traversing the line connecting their points in the
prior space.

In attempting to estimate how many distinct maps the generator can
produce, we generated a few batches of $32k$ maps each. We then examined
the pairwise distances among all the maps in each batch to try to find
maps that are ``close'' or identical to each other. We looked at
$L_{2}$ distance in pixels space, in the power spectra space and in the
space of the activations of the last conv-layer of the discriminator.
The maps determined as ``close'' in this way are still noticeably
distinct (visually and in the associated metric). If we are to use the
``Birthday Paradox'' test~\cite{bdayp}, this indicates that the number
of distinct maps the generator can produce is $\gg 1$B maps. It also
indicates that our generator does not suffer from mode collapse. Our
conjecture here is that one or a combination of multiple factors could
explain this result: (1) the convolution kernels might be the right
ansatz to describe the Gaussian fluctuations and the higher order
structures encoded in convergence maps, (2) the maps dataset contain
only one mode, so the normal prior, which has one mode, maps smoothly
and nicely to the mode of the convergence maps~\cite{bourgan}.

\section{Discussion and future work}%
\label{sec:discussion}
The idea of applying deep generative models techniques to emulating
scientific data and simulations has been gaining attention
recently~\cite{ravanbakhsh,calogan1,calogan2,mosser2017reconstruction}.
In this paper we have been able to reproduce maps of a particular
$\varLambda $CDM model with very high-precision in the derived summary
statistics, even though the network was not explicitly trained to
recover these distributions. Furthermore we have explored, and
reproduced the \emph{statistical variety} present in simulations, in
addition to the mean distribution. It remains an interesting question
for the applicability of these techniques to science whether the
increased precision achieved in this case, compared to other similar
work~\cite{webgan}, is due to the training procedure as outlined in
Sect.~\ref{sec:arch} or to a difference in the underlying data.

We have also studied the ability of the generator to generate novel maps, the
results shown in~Sect.~\ref{sec:generalize}. We have provided evidence that our
model has avoided `mode collapse' and produces distinct maps from the originals
that also smoothly vary when interpolating in the prior space.

Finally, the success of GANs in this work demonstrates that, unlike natural
images which are the focus of the deep learning community, scientific data come
with a suite of tools (statistical tests) which enable the verification and
improvement of the fidelity of generative networks. The lack of such tools for
natural images is a challenge to understanding the relative performance of the
different generative models architectures, loss functions and training
procedures~\cite{theis2015,Salimans2016}. So, as well as the impact within
these fields, scientific data with its suite of tools have the potential to
provide the grounds for benchmarking and improving on the state-of-the-art
generative models.

It is worth noting here that while one of the use cases of GANs is data
augmentation, we do not think the generated maps augment the original dataset
for statistical purposes. This is for the simple reason that the generator is a
fit to the data, sampling a fit does not generate statistically independent
samples. The current study is a proof-of-concept that highly over-parametrized
functions built out of convolutional neural networks and optimized in by means
of adversarial training in a GAN framework can be used to fit convergence maps
with high statistical fidelity. The real utility of GANs would come form the
open question of whether conditional generative
models~\cite{cgan,distill-feature-wise} can be used for emulating scientific
data and parameter inference. The problem is outlined below and is to be
addressed in future work.

We have also studied the ability of the generator to generate novel maps, the
results shown in~\ref{sec:generalize}. We have provided evidence that our model
has avoided `mode collapse' and produces distinct maps from the originals that
also smoothly vary when interpolating in the prior space.

Finally, the success of GANs in this work demonstrates that, unlike natural
images which are the focus of the deep learning community, scientific data come
with a suite of tools (statistical tests) which enable the verification and
improvement of the fidelity of generative networks. The lack of such tools for
natural images is one of the most serious challenges to understanding the
relative performance of the different generative models architectures, loss
functions and training procedures~\cite{theis2015,Salimans2016}. So, as well as
the impact within these fields, scientific data with its suite of tools have
the potential to provide the grounds for benchmarking and improving on the
state-of-the-art generative models.

It is worth noting here that while one of the use cases of GANs is data
augmentation, we do not think the generated maps augment the original dataset
for statistical purposes. This is for the simple reason that the generator is a
fit to the data, sampling a fit does not generate statistically independent
samples. The current study is a proof-of-concept that highly over-parametrized
functions built out of convolutional neural networks and optimized in by means
of adversarial training in a GAN framework can be used to fit convergence maps
with high statistical fidelity. The real utility of GANs would come form the
open question of whether conditional generative
models~\cite{cgan,distill-feature-wise} can be used for emulating scientific
data and parameter inference. The problem is outlined below and is to be
addressed in future work.

Without loss of generality, the generation of one random realization of a
science dataset (simulation or otherwise) can be posited as a black-box model
$S(\boldsymbol{\sigma }, r)$ where $\boldsymbol{\sigma }$ is a vector of the
physical model parameters and $r$ is a set of random numbers. The physical
model $S$ can be based on first principles or effective theories. Different
random seeds generate statistically independent mock data realizations of the
model parameters $\boldsymbol{\sigma }$. Such models are typically
computationally expensive to evaluate at many different points in the space of
parameters $\boldsymbol{\sigma }$.

In our minds, the most important next step to build on the foundation of this
paper and achieve an emulator of model $S$, is the ability of generative models
to generalize in the space of model parameters $\boldsymbol{\sigma }$ from
datasets at a finite number of points in the parameter space. More
specifically, \textit{can we use GANs to build parametric density estimators
$G(\boldsymbol{\sigma }, \boldsymbol{z})$ of the physical model
$S(\boldsymbol{\sigma }, r)$}? Such generalization rests on smoothness and
continuity of the response function of the physical model $S$ in the parameter
space, but such an assumption is the foundation of any surrogate modeling.
Advances in conditioning generative models~\cite{cgan,distill-feature-wise} is
a starting point to enable this goal. Future extensions of this work will seek
to to add this parameterization in order to enable the construction of robust
and computationally inexpensive emulators of cosmological models.

\section{Conclusion}\label{sec:conclusion}
We present here an application of Generative Adversarial Networks to the
creation of weak-lensing convergence maps. We demonstrate that the use
of neural networks for this purpose offers an extremely fast generator
and therefore considerable promise for cosmology applications. We are
able to obtain very high-fidelity generated images, reproducing the
power spectrum and higher-order statistics with unprecedented precision
for a neural network based approach. We have probed these generated maps
in terms of the correlations within the maps and the ability of the
network to generalize and create novel data. The successful application
of GANs to produce high-fidelity maps in this work provides an important
foundation to using these approaches as fast emulators for cosmology.

\small
\bibliographystyle{bibstyle}
\bibliography{bib}

\footnotesize

\section*{Declarations}
\subsection*{Data and code availability}
Code and example datasets with pretrained weights used to generate this
work are publicly available at
\url{https://github.com/MustafaMustafa/cosmoGAN}.\footnote{url date:
2019-03-15} The full dataset used in this work is available upon
request from the authors. We also provide the fully trained model which
was used for all results in this paper, how-to at link above.

\subsection*{Competing interests}
Biological authors declare that they have no competing interests. Two
artificial neural networks of this work: ``the generator'' and ``the
discriminator'' are however directly competing against each other,
effectively playing a zero-sum game.

\subsection*{Funding}
MM, DB and WB are supported by the U.S. Department of Energy,
Office of Science, Office of High Energy Physics, under contract No.
DEAC0205CH11231 through the National Energy Research Scientific
Computing Center and Computational Center for Excellence programs. ZL
was partially supported by the Scientific Discovery through Advanced
Computing (SciDAC) program funded by U.S.~Department of Energy Office
of Advanced Scientific Computing Research and the Office of High Energy
Physics.

\subsection*{Author Contributions} 
MM proposed the project, designed and carried out the experiments and
statistical analysis and prepared the first draft. DB  guided the cosmology and
statistical analysis, in addition to writing the introduction jointly with ZL
and co-writing the Results section. ZL provided guidance on cosmology
simulations and emulators in addition to co-writing the Discussion section. WB
contributed to interpreting the Results and Discussions of future work in
addition to reviewing the manuscripts. RA contributed to discussions of the
deep learning aspects of the project and reviewing the manuscript. JK produced
the simulated convergence maps. All authors read and approved the final
manuscript.

\subsection*{Acknowledgements} 
MM would like to thank Evan Racah for his invaluable deep learning
related discussions throughout this project. MM would also like to
thank Taehoon Kim for providing an open-sourced TensorFlow
implementation of DCGAN which was used in early evaluations of this
project. We would also like to thank Ross Girshick for helpful
suggestions. No network was hurt during the adversarial training
performed in this work.
\end{document}